\documentstyle[prl,aps,floats,epsf,psfig,colordvi]{revtex}
\newcommand{\alfs}{\mbox{$\alpha_s$}}
\newcommand{\mztwo}{\mbox{$M_Z^2$}}
\def\np#1#2#3   {{ Nucl. Phys.} {\bf#1}, #2 (#3). }
\def\pcps#1#2#3 {{ Proc. Cam. Phil. Soc.} {\bf#1}, #2 (#3). }
\def\pl#1#2#3   {{ Phys. Lett.} {\bf#1}, #2 (#3). }
\def\plc#1#2#3   {{ Phys. Lett.} {\bf#1}, #2 (#3); }
\def\prep#1#2#3 {{ Phys. Rep.} {\bf#1}, #2 (#3). }
\def\prev#1#2#3 {{ Phys. Rev.} {\bf#1}, #2 (#3). }
\def\prl#1#2#3  {{ Phys. Rev. Lett.} {\bf#1}, #2 (#3). }
\def\prs#1#2#3  {{ Proc. Roy. Soc.} {\bf#1}, #2 (#3). }
\def\ptp#1#2#3  {{ Prog. Th. Phys.} {\bf#1}, #2 (#3). }
\def\rmp#1#2#3  {{ Rev. Mod. Phys.} {\bf#1}, #2 (#3). }
\def\rpp#1#2#3  {{ Rep. Prog. Phys.} {\bf#1}, #2 (#3). }
\def\zp#1#2#3   {{ Z. Phys.} {\bf#1}, #2 (#3). }
\def\epj#1#2#3   {{ Eur. Phys. Jour.} {\bf#1}, #2 (#3). }

\begin{document}

\wideabs{
\title{Parton  Distributions, $d/u$, and Higher Twists Effects at High $X$}

\author{ U.~K.~Yang and A.~Bodek}

\address{Department of Physics and astronomy,
University of Rochester, Rochester, NY 14627 }

\maketitle
\begin{abstract}
A re-analysis of the  NMC and SLAC data leads to a
great improvement in our  knowledge of the valence
$d$ and $u$ parton distribution functions (PDF's) at high $x$. 
Standard parton distributions with our modifications
are in good agreement with QCD predictions for $d/u$
at $x=1$, and with the CDHSW $\nu p$ and $\overline{\nu} p$
data, the HERA charged current cross section data,
the collider high-$P_t$ jet data, and the CDF
$W$ asymmetry data. With the inclusion of target mass
and  higher twist corrections, the modified PDF's
also describe all DIS data up to $x = 0.98$ and down to
$Q^2 = 1$ GeV$^2$.
\end{abstract}
\vspace{-0.4in}
\twocolumn
}


Recent work on parton distributions functions (PDF's)
in the nucleon has focussed
on probing the sea and gluon distribution at small $x$. The valence 
quarks distribution has been thought to be relatively well understood.
However, the precise knowledge of the $u$ and $d$ quark distribution
at high $x$ is very important at collider energies in searches for signals
for new physics at high $Q^2$.
In addition, the value of $d/u$ as $x \rightarrow 1$ is of theoretical
interest.
Recently, a proposed CTEQ toy model~\cite{toy} included
 the possibility of an additional contribution to the $u$ quark distribution
(beyond $x>0.75$) as an explanation for both the initial HERA high
 $Q^2$ anomaly~\cite{highQ2},
and for the jet excess at high-$P_t$  at CDF~\cite{CDFjet}. In this
Letter we conclude that a re-analysis of data from
NMC and SLAC leads to a great improvement in our  knowledge
of PDF's at high $x$, and rules out such toy models.

\begin{figure}[t]
\centerline{\psfig{figure=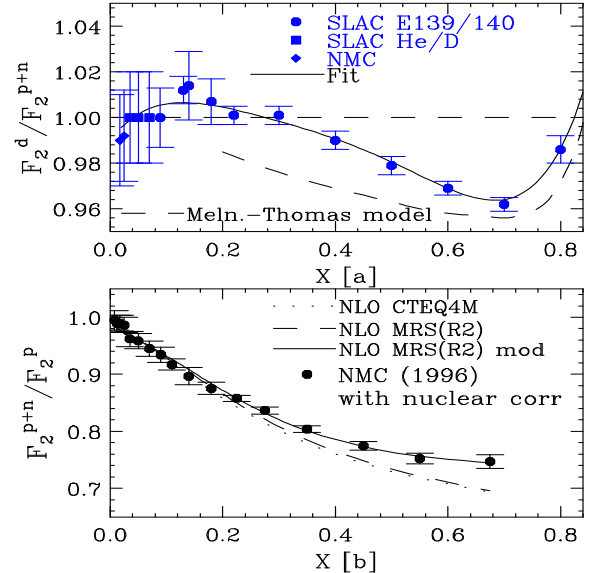,width=3.0in,height=3.0in}}
\caption{(a) The total correction for nuclear 
effects (binding and Fermi motion) in the deuteron,
 $F_2^d/F_2^{n+p}$, as a function of $x$, extracted from fits to
the nuclear dependence of SLAC $F_2$ electron scattering
data (compared to theoretical model~[6]). ~(b) Comparison of NMC $F_2^{n+p}/F_2^p$ (corrected for nuclear
effects) and the prediction in NLO using the  MRS(R2) 
PDF
with and without our proposed modification to the $d/u$ ratio.}
\label{fig:f2dp}
\end{figure}

Information about valence quarks originates from the proton and neutron
structure function data. The $u$ valence quark distribution at high $x$
is relatively well
constrained by the proton structure function $F_2^p$.
However, the neutron structure function $F_2^n$, which is sensitive 
to the $d$ valence quark at high $x$,
is actually extracted from deuteron data.
 Therefore, there is an uncertainty in the $d$ valence quark 
distribution from the corrections for nuclear binding effects in the deuteron. 
In past extractions of $F_2^n$ from deuteron data,
only Fermi motion corrections
were considered, and other 
binding effects were assumed to be negligible.
Recently, the corrections for nuclear binding effects in the deuteron,
$F_2^d/F_2^{n+p}$, have been extracted empirically from
fits to the nuclear dependence
of electron scattering data from SLAC experiments
E139/140~\cite{GOMEZ}.
The empirical extraction uses a
model proposed by Frankfurt and Strikman~\cite{Frankfurt}, 
in which all binding effects 
in the deuteron and heavy nuclear targets
are assumed to scale with the nuclear density.
The total correction for nuclear binding effects in the deuteron
(shown in Fig. \ref{fig:f2dp}(a)), 
is in a direction which is opposite
to what is expected from the previous models which included only 
the Fermi motion effects. 
The suprisingly large correction extracted in this empirical way
maybe controversial, but is smaller than the recent theoretical prediction
~\cite{duSLAC} (dashed line in Fig. \ref{fig:f2dp}(a))

\begin{figure}[ht]
\centerline{\psfig{figure=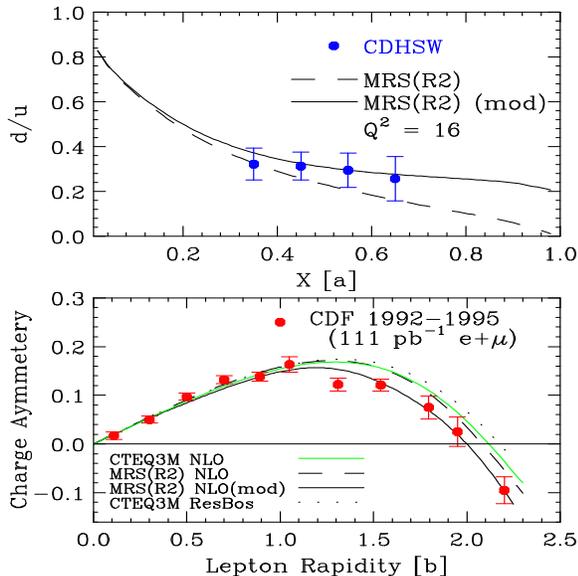,width=3.0in,height=3.0in}}
\caption{(a) The $d/u$ distributions at $Q^2$=$16$ GeV$^2$ 
as a function of $x$ for the standard and modified MRS(R2) PDF
compared to the CDHSW data. 
(b) Comparison of the CDF $W$ asymmetry data with NLO standard
CTEQ3M, MRS(R2), and modified MRS(R2) as a function of the lepton rapidity.
The standard CTEQ3M with 
a resummation calculation is also shown 
for comparison.}
\label{fig:dou}
\end{figure}

The ratio $F_2^d/F_2^p$ is directly related to $d/u$. In leading order QCD,
 $2F_2^d/F_2^p -1 \simeq (1+4d/u)/(4+d/u)$ at high $x$.
  We perform a next-to-leading order (NLO) analysis on the precise
  NMC $F_2^d/F_2^p$ data~\cite{NMCf2dp} to extract $d/u$
  as a function of $x$.
 We extract the ratio $F_2^{p+n}/F_2^p$
 by applying the nuclear binding correction
 $F_2^d/F_2^{n+p}$  to the $F_2^d/F_2^p$ data.
 As shown in Fig. \ref{fig:f2dp}(b), the
 standard PDF's~\cite{MRSR2,CTEQ3M} do
 not describe the extracted $F_2^{p+n}/F_2^p$.
 Since the $u$ distribution is relatively well constrained,
 we find a correction term to  $d/u$ in the standard PDF's
 (as a function of $x$), by  varying only the $d$ distribution to fit the data.
 The correction term is  parametrized  
 as a simple quadratic form, $\delta (d/u) = (0.1\pm0.01)(x+1)x$
 for the MRS(R2) PDF,
 where the corrected $d/u$ ratio
 is $(d/u)' = (d/u) + \delta (d/u)$.
 Based on this correction,
 we obtain a  MRS(R2)-modified PDF as shown in Fig \ref{fig:dou}(a).
 The correction to other PDF's such as CTEQ3M/4M is similar.
 Note that since the $d$ quark level is small at large $x$,
 all the sum rules are easily satisfied with a very minute change at low $x$.
 The NMC data, when corrected for nuclear binding effects
in the deuteron, clearly indicate that $d/u$ in the
 standard PDF's is significantly underestimated 
 at high $x$ as shown in Fig. \ref{fig:dou}(a).
 It also shows that the modified $d/u$ ratio 
 approaches
 $0.2\pm0.02$ as $x \rightarrow 1$, in agreement with a QCD
 prediction~\cite{Farrar}. In contract, if the
deuteron data is only corrected for Fermi motion effects (as
was mistakenly done in the past) both the $d/u$ from data and the $d/u$
in the standard PDF's fits
approach $0$ as $x \rightarrow 1$.
 Figure~\ref{fig:dou}(a) shows that  $d/u$ values 
extracted from CDHSW~\cite{du_cdhsw} $\nu p$/
$\overline{\nu} p$ data (which are free from nuclear effects) 
also favor the modified PDF's at high $x$.

\begin{figure}
\centerline{\psfig{figure=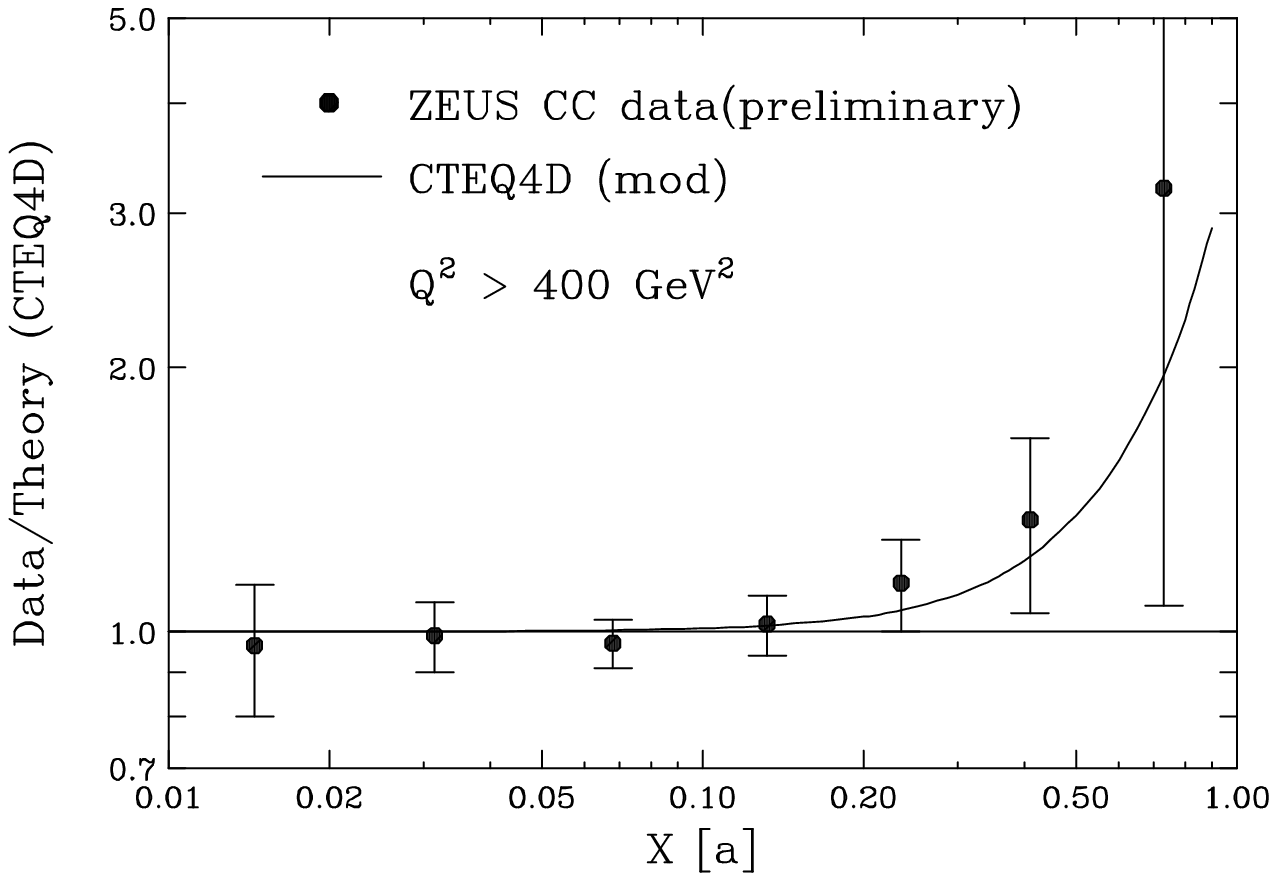,width=3.0in,height=1.5in}}
\centerline{\psfig{figure=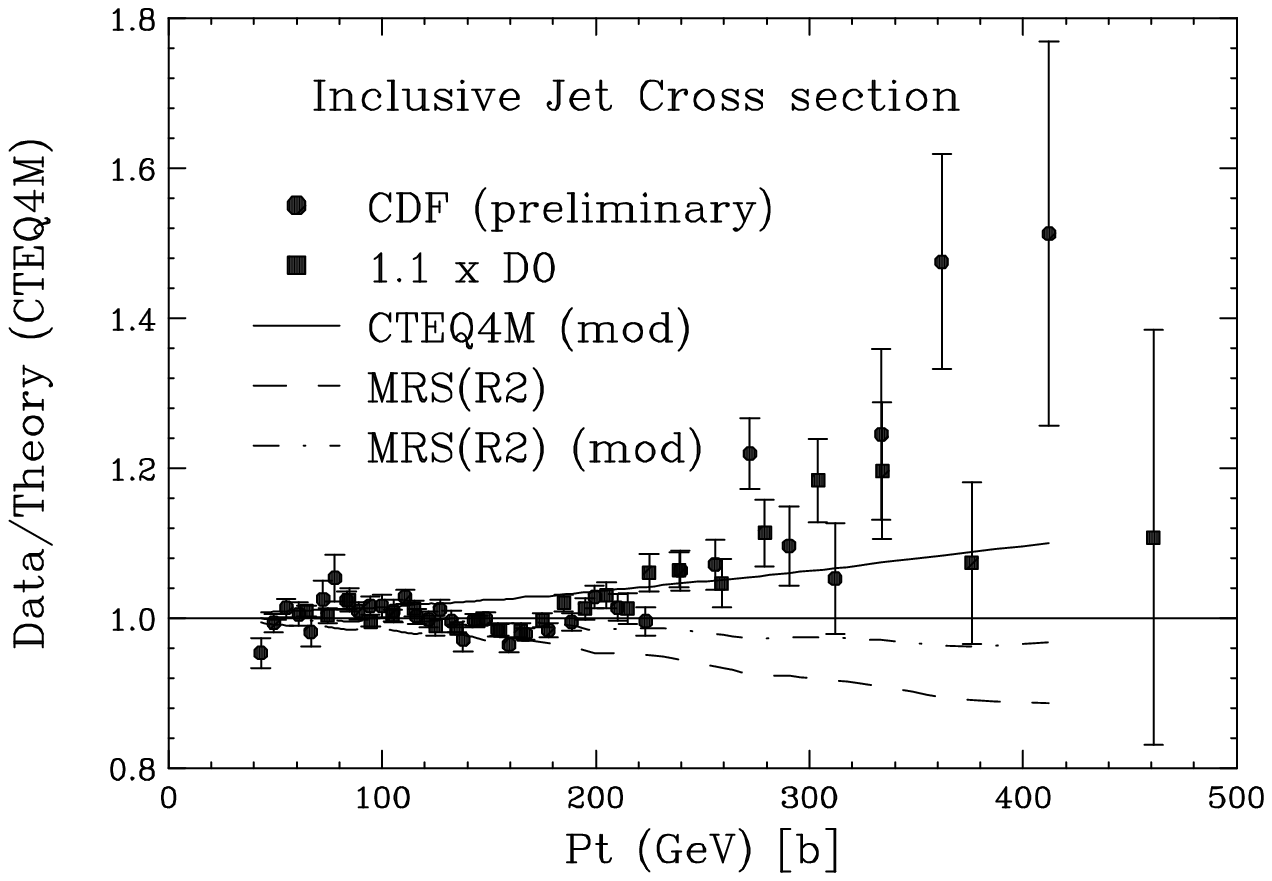,width=3.0in,height=1.5in}}

\caption{ (a) The HERA charged current cross section data and 
(b) the CDF and D0 
inclusive jet cross section data are compared 
with both standard and modified PDF's.}
\label{fig:highq2}
\end{figure}

Information (which is not affected by the corrections
for nuclear effects in the deuteron)
on $d/u$ can be also extracted from $W$ production data in hadron
colliders.
Figure~\ref{fig:dou}(b) shows that the predicted $W$ asymmetry calculated
with the DYRAD NLO QCD program
using our modified PDF is
in much better agreement with recent CDF data~\cite{Wasym} at large
rapidity than standard PDF's.
When the modified PDF at $Q^2$=$16$ GeV$^2$ is evolved to $Q^2$=$10^4$ GeV$^2$
using the NLO QCD evolution, we find that
the modified $d$ distribution at $x=0.5$ is increased by about 40 \% 
in comparison to the standard $d$ distribution.
The modified PDF's have a significant impact
on the 
charged current cross sections~\cite{zeushighq2}
in the HERA high $Q^2$ region, shown in 
Fig.~\ref{fig:highq2}(a), because 
the charged current scattering with positrons is on $d$ quark only.
Figure~\ref{fig:highq2}(b) shows that
the modified PDF's
also lead to an increase of 10\% in the
production rate
of very high $P_T$ jets~\cite{jet} in hadron colliders. 

Since all the standard PDF's, including our modified versions, are
fit to data with $x$ less than 0.75, we now
investigate the validity of the modified MRS(R2) at very high
$x$  by comparing to  $F_2^p$ data at SLAC.
Although the SLAC data at very high $x$ are at reasonable values
of $Q^2$ $(7<Q^2<31$ GeV$^2)$, they are in a region in which
 non-perturbative effects such as target mass
 and higher twist are very large.
We use the Georgi-Politzer calculation~\cite{GPtm}
for the target mass corrections (TM). These involve
using the scaling variable 
$\xi=2x/(1+\sqrt{1+4M^2x^2/Q^2})$ instead of $x$.
Since a complete calculation of higher twist
effects is not available, the very low $Q^2$ data are
used to obtain information on the size of these terms.

We use two approaches in our study of the higher
twist effects:
an empirical method, and the renormalon model. 
In the empirical approach, the higher twist contribution is evaluated
by adding a term $h(x)/Q^2$ to the perturbative QCD (pQCD) prediction 
of the structure function (including target mass effects).
 The $x$ dependence of the higher
twist coefficients $h(x)$ is fitted to the global 
deep-inelastic scattering (DIS) $F_2$ 
(SLAC, BCDMS, and NMC) data~\cite{SLACF2,BCDMSF2,NMCF2} in the 
kinematic region 
($0.1<x<0.75$, $1.25<Q^2<260$ GeV$^2$)
with the following form; 
$F_2 = F_2^{pQCD+TM}(1+h(x)/Q^2)f(x)$. 
Here $f(x)$ is a floating factor to investigate possible
$x$ dependent corrections to our modified PDF.
A functional form, $a(\frac{x^b}{1-x}-c)$ 
for $h(x)$ is used in the higher twist fit to
estimate the size of the higher twist terms above $x=0.75$.
The SLAC and BCDMS data are normalized to the NMC data.
In the case of the BCDMS data, a systematic error shift $\lambda$
(in standard deviation units) is allowed
to account for the correlated point-to-point  systematic errors.
The empirical higher twist fits with the modified NLO MRS(R2) pQCD
prediction with TM
have been performed simultaneously
on the proton and deuteron $F_2$ data with 11 free parameters
(two relative normalizations and three parameters for $h(x)$ per target
and the BCDMS $\lambda$).
We find that empirical higher twist fit describes the data well
($\chi^2/d.o.f.=843/805$), and the higher twist contributions
in the  proton and deuteron are similar.
The magnitude is  almost half of the size from
a previous analysis of SLAC/BCDMS
data~\cite{Virchaux},
because that analysis was based on \alfs(\mztwo) $=0.113$,
while \alfs(\mztwo) $=0.120$ in the MRS(R2) PDF,
which is close to the current world average.

\begin{figure}
\centerline{\psfig{figure=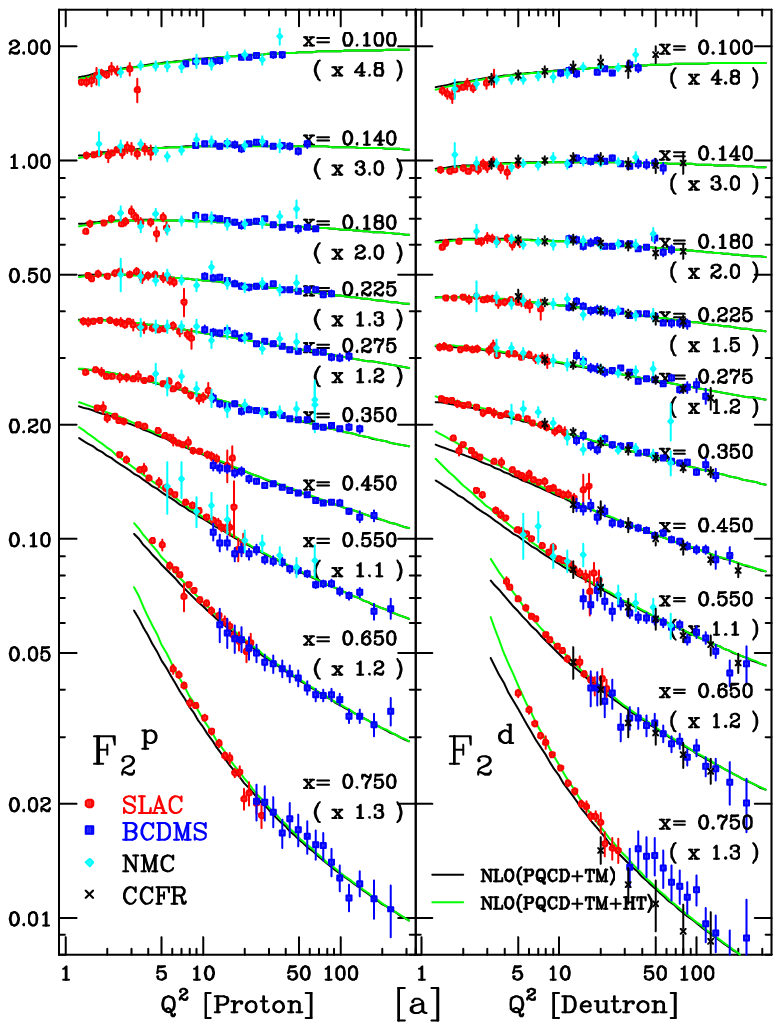,width=3.1in,height=4.1in}}
\centerline{\psfig{figure=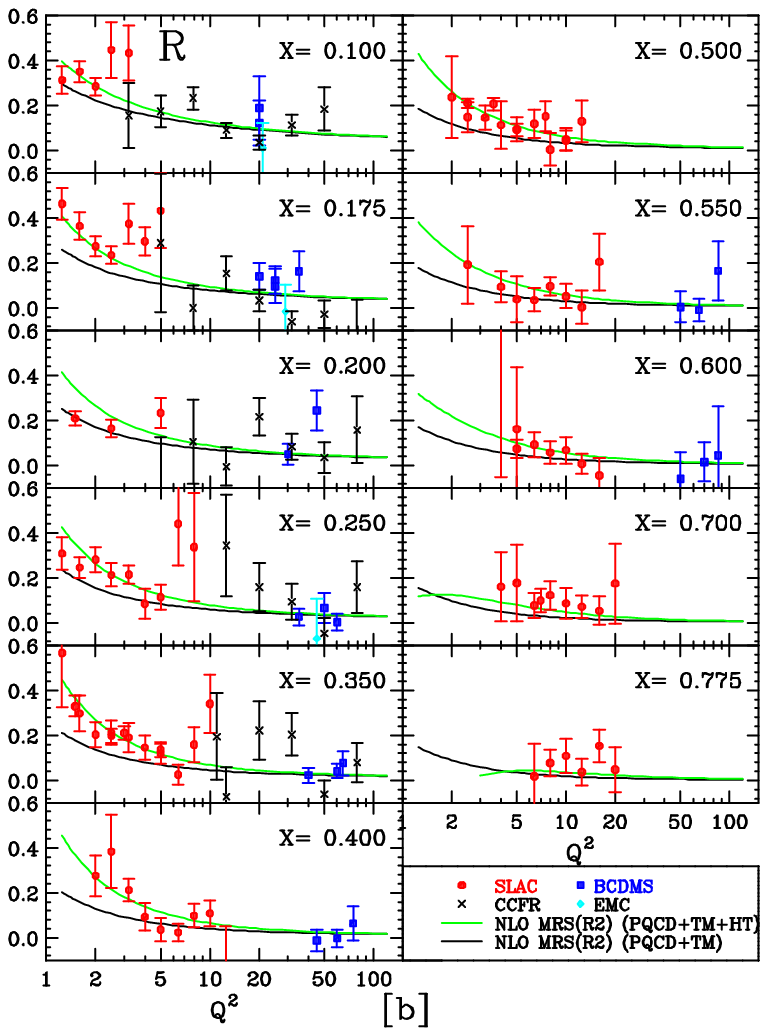,width=3.1in,height=4.1in}}
\caption{The description of higher twist fit using the renormalon model
with the modified NLO MRS(R2) PDF. The CCFR neutrino data is also shown 
for comparison.
(a) Comparison of $F_2$ and NLO prediction
with and without higher twist contributions. 
(b) Comparison of $R$ and NLO prediction
with and without 
the renormalon higher twist contributions.}
\label{fig:disht}
\end{figure}
In the renormalon model approach~\cite{renormalon}, the 
model predicts the complete $x$ dependence of the higher twist
contributions 
 to  $F_2$, $2xF_1$, and $xF_3$, with only two unknown parameters
$A_2$ and $A_4$. 
We extract the $A_2$ and $A_4$ parameters, which
determine the overall level of the  $1/Q^2$
and $1/Q^4$ terms by fitting to the global
data set for $F_2$ and $R [= F_2(1+4Mx^2/Q^2)/2xF_1 - 1]$.
The values of $A_2$ and $A_4$ for the proton and deuteron
are same in this model.
The $x$ dependence of $2xF_1$ differs from that of $F_2$ but is the 
same as that
of $xF_3$ within a power correction of $1/Q^2$. 
Our fits  can also be used to estimate the size of the higher twist effects
in $xF_3$ [e.g. the Gross-Llewellyn Smith (GLS) sum rule].
The higher twist fit in this approach has employed the same procedure 
as the empirical method.  Figure~\ref{fig:disht} shows that the model yields
description of the $x$ dependence of higher twist terms in
 both $F_2$ and $R$
with just the two free parameters ($\chi^2/d.o.f.=1470/928$).
The CCFR neutrino data~\cite{CCFRF2}
is shown for comparison though it is not used in the fit.
The extracted values of $A_2$ are $-0.093 \pm 0.005$
 and $-0.101 \pm 0.005$, for proton and deuteron,
 respectively. The contribution of $A_4$ is found  to be negligible. 
We find that the floating factor $f(x)$ for the deuteron deviates from 1
and is also bigger than
that for the proton, unless the modified MRS(R2) PDF is used.
This reflects our earlier conclusion
that the standard $d$ distribution is underestimated at high $x$ region.
As in the empirical fit, the extracted $A_2$ value 
is half of the previous estimated value~\cite{renormalon} 
of $A_2$ based on SLAC/BCDMS [\alfs(\mztwo) $=0.113$] analysis~\cite{note}. 
Since both of these approaches yield a reasonable description for the higher
twist effects, we proceed to compare the predictions of the modified
PDF's (including target mass and renormalon higher twist corrections) to the
SLAC proton $F_2$ data at very high $x$ ($0.7<x<1$). 


\begin{figure}[t]
\centerline{\psfig{figure=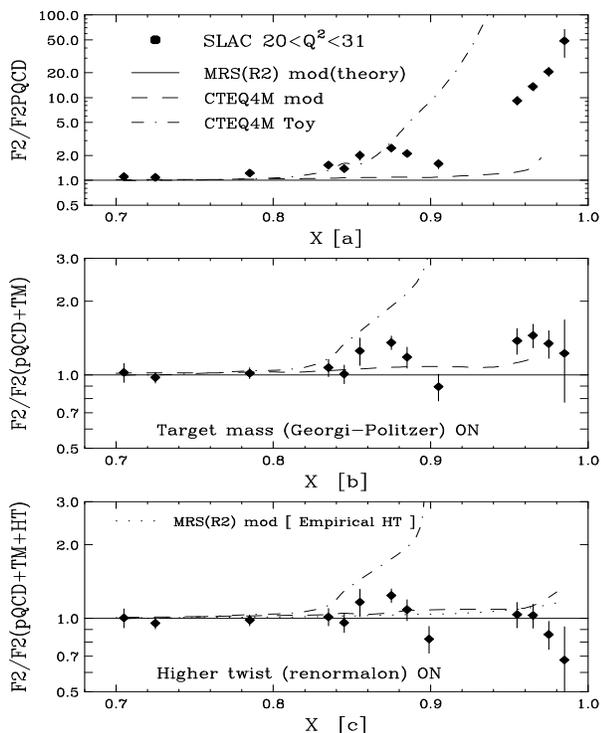,width= 3.1in,height=3.8in}}
\caption{Comparison of SLAC $F_2^p$ data
with the predictions of the modified MRS(R2), CTEQ4M and the CTEQ toy model at high $x$
and higher $Q^2$ ($20<Q^2<31$ GeV$^2$). ~~~(a) Ratio to pQCD, (b) ratio to pQCD 
with TM effects,
and (c) ratio to pQCD with TM and higher twist effects}
\label{fig:highx_highq2}
\end{figure}

There is a wealth of SLAC data~\cite{SLACres} in the region up to $x=0.98$ 
and intermediate $Q^2$ $(7<Q^2<31$ GeV$^2)$. 
Previous PDF fits have not used these data. 
We use the estimate of the higher twist effects 
from the models, based on the data (below $x<0.75$) described above.
Note that the data for $x>0.75$ is in the DIS region,
and the data for $x>0.9$ is the resonance region.
It is worthwhile to investigate the resonance region also because
from duality arguments~\cite{Bloom} it is expected  
that the average behavior of the resonances and elastic peak
should follows the DIS scaling limit curve.
Figure~\ref{fig:highx_highq2} shows the ratio of the SLAC data to the predictions
of the modified MRS(R2) at relatively large $Q^2$ ($21<Q^2<30$ GeV$^2$)
where the elastic contribution is negligible.
With the inclusion of target mass and the renormalon higher twist
effects, the very high $x$ data from SLAC is remarkably well described 
by the modified MRS(R2) up to $x=0.98$.  
The good description
of the data by the modified MRS(R2) is also achieved 
using the empirical estimate [$h(x)/Q^2$] of higher twist effects
as shown in Fig.~\ref{fig:highx_highq2}(c).
Figure~\ref{fig:highx_highq2} also shows that the CTEQ Toy model
(with an additional 0.5\%
component of $u$ quarks beyond $x>0.75$) overestimates the SLAC data 
by a factor of three at $x = 0.9$ (DIS region).
From these comparisons, we find that the SLAC $F_2$ data do not support
the CTEQ Toy model
which proposed an additional $u$ quark contribution
at high $x$ as an explanation of the initial HERA high $Q^2$ anomaly
and the CDF high-$P_t$ jet excess. As indicated in
Fig.~\ref{fig:highx_highq2}(c), the uncertainties in the PDF's at
high $x$ are small. The difference between
CTEQ4M and MRS(R2) (with our $d/u$ modifications) is
an estimate of the errors.

In conclusion, we find that nuclear binding effects in the deuteron
play a significant role in our understanding of $d/u$
at high $x$.
With the inclusion of target mass
and  higher twist corrections, the modified PDF's
also describe all DIS data up to $x = 0.98$ and down to
$Q^2 = 1$ GeV$^2$.
The modified PDF's with our $d/u$ correction
are in good agreement with the prediction of QCD at $x=1$, and with
the CDHSW $\nu p$
and $\overline{\nu} p$
data, the HERA CC cross section data,
the collider high-$P_t$ jet data, and with the
CDF $W$ asymmetry data.
A next-to-next leading order (NNLO) analysis~\cite{note} of $R$ indicates  that
the higher twist effects extracted in the NLO fit 
at low $Q^2$ may originate from the missing NNLO terms.


\end{document}